\documentclass[12pt,a4paper,reqno]{amsart}

\pdfoutput=1

\usepackage{graphicx}
\usepackage{amssymb}
\usepackage{amsfonts}
\usepackage{amsmath}
\usepackage[normalsize]{subfigure}
\usepackage[headings]{fullpage}
\usepackage[dvips,colorlinks,bookmarksopen,bookmarksnumbered,citecolor=red,urlcolor=red]{hyperref}

\newcommand{\figName}[1]{#1}

\newcommand{\mbf}[1]{{\boldsymbol #1}}
\newcommand{\Sref}[1]{Section~\ref{#1}}
\newcommand{\Fref}[1]{Figure~\ref{#1}}
\newcommand{\Eref}[1]{Eq.~(\ref{#1})}

\newcommand{\set}[1]{{\mathbb #1}}
\newcommand{\EX}[1]{{\mathbb E} \left[ #1 \right]}
\newcommand{\mean}[1]{\mu_{#1}}
\newcommand{\meanv}[1]{\mbf{\mu}_{#1}}
\newcommand{\pme}{\mathcal{P}} 
\newcommand{\rl}{\theta} 

\newcommand{\de}[1]{\,{\mathrm d}#1} 
\newcommand{\corr}[2]{R_{#1#2}}
\newcommand{\acorr}[1]{R_{#1}}
\newcommand{\acorrm}[1]{\mbf{R}_{#1}}

\newcommand{\var}[1]{\sigma_{#1}^2}
\newcommand{\phs}[1]{^{(#1)}}

\newcommand{\tppF}[1]{S\phs{#1}_2}

\newcommand{\h}{_{h}}
\renewcommand{\u}{^{u}}
\renewcommand{\t}{^{\tau}}
\newcommand{\trn}{{\sf ^T}}
\newcommand{\SFE}{\rm{SFE}}
\newcommand{\KLE}{\rm{KLE}}
\newcommand{\HS}{\rm{HS}}

\DeclareMathOperator*{\stat}{\mathrm{stat}}

\newcommand{\half}{\mbox{$\frac{1}{2}$}}

\title[Stochastic modeling of chaotic masonry $\ldots$]{Stochastic
  modeling of chaotic masonry via mesostructural characterization}

\author{M. Lombardo}
\address{%
Department of Civil Engineering, University of Messina\\
C.da di Dio, 98166 Messina, Italy}
\email{mlombardo@ingegneria.unime.it}

\author{J.~Zeman}
\address{%
Department of Mechanics, Faculty of Civil Engineering, Czech Technical
University in Prague\\
Th\' akurova 7, 166 29 Prague 6, Czech Republic
}
\email{zemanj@cml.fsv.cvut.cz}
\urladdr{http://mech.fsv.cvut.cz/~zemanj}

\author{M.~\v{S}ejnoha}
\address{%
Department of Mechanics, Faculty of Civil Engineering, Czech Technical
University in Prague, Th\' akurova 7, 166 29 Prague 6, Czech Republic}
\address{
Centre for Integrated Design of Advanced Structures, Th\' akurova 7,
166 29 Prague 6, Czech Republic}
\email{sejnom@fsv.cvut.cz}
\urladdr{http://mech.fsv.cvut.cz/~sejnom}

\author{G. Falsone}
\address{%
Department of Civil Engineering, University of Messina\\
C.da di Dio, 98166 Messina, Italy}
\email{gfalsone@ingegneria.unime.it}

\begin{document}

\begin{abstract}
The purpose of this study is to explore three numerical approaches to
the elastic homogenization of disordered masonry structures with
moderate meso/macro-lengthscale ratio. The methods investigated
include a representative of perturbation methods, the
Karhunen-Lo\`{e}ve expansion technique coupled with Monte-Carlo
simulations and a solver based on the Hashin-Shtrikman variational
principles. In all cases, parameters of the underlying random field of
material properties are directly derived from image analysis of a
real-world structure. Added value as well as limitations of individual
schemes are illustrated by a case study of an irregular masonry panel.
\end{abstract}

\maketitle

\section*{Keywords}
irregular masonry structures; stochastic homogenization; two-point
statistics; improved perturbation methods; Karhunen-Lo\`{e}ve
expansion; Hashin-Shtrikman variational principles

\section{Introduction}\label{Introduction}
The last decade has witnessed rapid advances in modelling and
simulation of masonry structures, mainly in connection with
reconstruction and rehabilitation of historical monuments. The major
impetus for these developments came from a seminal contribution by
Anthoine~\cite{Anthoine:1995:DIPE}, which clearly demonstrated that
the complex overall behavior of elastic regular masonry can be
systematically addressed in the framework of homogenization theory for
periodic heterogeneous media. These results were subsequently extended
towards tools capable of predicting non-linear response of masonry
structures with deterministic geometry and sufficiently small ratio
between the characteristic size of masonry bond~(mesoscale) and the
structural level~(macroscale),
see~\cite{Lourenco:2007:AMS,Massart:2007:EMS} for an extensive
overview of the field. Both these assumptions, however, may show to be
inadequate for historical structures, where the spatial distribution
of individual constituents is random rather then deterministic and the
typical size of a block may well become comparable to the macroscopic
lengthscale.

When adopting certain simplifying assumptions, a vast body of
approaches is currently available for the treatment of irregular
masonry as a random heterogeneous material. Under the hypothesis of
widely separated lengthscales, the well-established tools of
\emph{stochastic continuum micromechanics} can be adopted, in which
the inhomogeneous body in question is replaced with a homogeneous
equivalent with properties determined from analysis of a
representative volume element, locally replacing the mesoscale level,
see monographs~\cite{Buryachenko:2007:MHM,Torq02} for up-to-date
reviews. To the authors' best knowledge, the only masonry-related
study available in this field was presented by
\v{S}ejnoha~et~al.~\cite{Sejnoha:2004:HRMS} in the framework of
stochastic re-formulation of Hashin-Shtrikman variational principles
due to Willis~\cite{Willis:1977:BSC}. Additionally, the notion of a
\emph{stochastic representative element} can be adopted, based either
on matching spatial statistics as originally proposed by Povirk
in~\cite{Povirk:1995:IMI} and subsequently applied to masonry
structures in~\cite{Sejnoha:2004:HRMS,Zeman:2007:FRM}, or deduced from
the convergence of apparent macroscopic properties. The latter concept
was proposed by Huet~\cite{Huet:1990:AVC} in the deterministic
setting, extended by Sab~\cite{Sab:1992:HSRM} to random media and
implemented for historical masonry structures by Cluni and
Gusella~\cite{Cluni:2004:HNP,Gusella:2006:RFH}.

Complementary, systems with random material properties quantified by
\emph{random variables} and deterministic (or slightly perturbed)
geometry of the representative volume can be treated employing
numerical techniques of stochastic mechanics such as Monte-Carlo
simulations~\cite{Kaminski:1996:HER}, perturbation-based
methods~\cite{Kaminski:2000:PBSFEM,Sakata:2008:3DSA} or approaches
based on an empirical probability distribution
function~\cite{Sakata:2008:KBA}; see also~\cite{Kaminski:2004:CMCM}
for a systematic overview. Most generally, uncertainties in spatial
distribution and material properties of individual phases can be
jointly characterized when resorting to random field
description~\cite{Vanm98}. Under suitable assumptions on the
underlying random field, a rigorous homogenization theory available
in~\cite{Bensoussan:1978:AAPS,Jikov:1994:HDE} was used to construct
efficient stochastic homogenization solvers, based on spectral
collocation methods~\cite{Jardak:2004:SSH} or Fourier-Galerkin
approaches~\cite{Xu:2005:SCM,Xu:2006:CSH}. Recently, these methods
were extended by Xu~\cite{Xu:2007:MSFEM} to treat heterogeneous media
with small but finite lengthscale contrast.

Finally, as the macro/meso scale ratio further increases, the rapidly
developing tools of Stochastic Finite Element~(\SFE)
methods~\cite{Babuska:2005:SEB,GhaSpa91a,Matthies:2005:GMLN,SudDer00}
become applicable for the assessment of overall response. It should be
emphasized that even though the random field description definitely
offers a more general framework than the alternative schemes, its
major weakness is that the random field is often introduced without a
clear link with the heterogeneous mesostructure,
see~\cite{Charmpis:2007:NLM} for a lucid discussion on the
topic. Moreover, the application of the random field/variable
paradigms to the simulation of masonry structure seems to be currently
missing; the only related works the authors are aware of is a recent
contribution of Spence et al.~\cite{Spence:2008:PMS} dealing with
mesostructure generation of irregular masonry walls.

In this contribution, three numerical approaches to the determination
of the overall response of elastic masonry with comparable macro- and
meso-lengthscales are presented. A unifying feature is the description
of mechanical properties in the form of a random field with the
second-order statistics consistently derived from image analysis of
the investigated structure. In \Sref{Microstructural_statistics}, this
procedure is briefly summarized following the exposition of Falsone
and Lombardo~\cite{Falsone:2007:SRMP}. The second level of
representation involves the determination of basic statistics related
to the response of a finite size heterogeneous masonry structure. In
particular, the improved perturbation method is introduced first
in~\Sref{Perturbative_approach}, followed in~\Sref{KL-expansion} by
the Monte-Carlo approach with individual realizations of the random
field generated using the Karhunen-Lo\`{e}ve
expansion. \Sref{Hashin-Shtrikman} is concerned with the application
of the Hashin-Shtrikman variational principles, coupled with the
Finite Element discretization to allow for the treatment of
finite-size bodies.  In~\Sref{Numerical_Example}, the results obtained
with the selected methods are mutually compared on the basis of
elastic analysis of an irregular masonry panel. Finally,
\Sref{Conclusions} introduces possible extensions and refinements of
the studied approaches.

In the following text, the Voigt representation of symmetric tensorial
quantities is systematically employed, see
e.g.~\cite{Bittnar:1996:NMM}. In particular, $a$, $\mbf{a}$ and
$\mbf{A}$ denote a scalar value, a vector or a matrix representation
of a second-order tensor and a matrix representation of a fourth-order
tensor, respectively.

\section{Probabilistic characterization of material property via mesostructural statistics}\label{Microstructural_statistics}

Before getting to the heart of the matter, we begin by summarizing
essential terminology related to the theory of random
fields~\cite{Vanm98}. Given a complete probability space $\{ \Theta,
\mathcal{F}, \mathcal{P} \}$ with sample space $\Theta$,
$\sigma$-algebra $\mathcal F$ on $\Theta$ and probability measure
$\mathcal P$ on $\mathcal F$, a scalar random field $H$ defined on an
open set $\Omega \subset \set{R}^d$ is a mapping
\begin{equation}
H : \Theta \times \Omega \rightarrow \set{R},
\end{equation}
such that, for every $\mbf{x} \in \Omega$, $H(\mbf{x}; \theta)$ is a
random variable with respect to the triple $\{ \Theta, \mathcal{F},
\mathcal{P} \}$. The mean of a random field is then given as
\begin{equation}\label{eq:mean}
\mean{H}( \mbf{x} )
=
\EX{ H(\mbf{x}; \rl )}
=
\int_\Theta
H(\mbf{x}; \rl )
\de \pme(\rl),
\end{equation}
for any $\mbf{x} \in \Omega$, whereas the covariance of two random
fields $H$ and $G$ is defined by
\begin{equation}\label{eq:corr}
\corr{H}{G}( \mbf{x}, \mbf{x}' )
=
\EX{%
\left( H(\mbf{x}; \rl) - \mean{H}(\mbf{x}) \right)
\left( G(\mbf{x}'; \rl) - \mean{G}(\mbf{x}') \right)
},
\end{equation}
with the symbol $\acorr{H}( \mbf{x}, \mbf{x}' ) = \corr{H}{H}(
\mbf{x}, \mbf{x}' )$ reserved for the autocovariance, reducing to a
variance for $\mbf{x} = \mbf{x}'$:
\begin{equation}
\var{H}(\mbf{x} )
=
\EX{%
\left( H(\mbf{x}; \rl ) - \mean{H}(\mbf{x}) \right)^2
}.
\end{equation}
A random field $H(\mbf{x}; \rl)$ is said to be homogeneous if all
its joint probability distribution functions~(PDFs) remain
invariant under the translation of the coordinate system, leading
to substantial simplification of the considered statistics:
\begin{eqnarray}
\mean{H}(\mbf{x})=\mean{H}=\mbox{const}, &
\var{H}(\mbf{x})=\var{H}=\mbox{const}, &
\acorr{H}(\mbf{x},\mbf{x}' ) = \acorr{H}(\mbf{x} - \mbf{x}' ).
\end{eqnarray}
Assuming that the autocovariance function can be well-approximated by
an exponential function, the correlation length $\lambda_H$ is defined
by means of inequality:
\begin{equation}
\forall 
\|\mbf{x} - \mbf{x}' \| \geq \lambda_H
:
\acorr{H}(\mbf{x} - \mbf{x}' ) \leq \frac{\var{H}}{\exp(1)},
\end{equation}
hence quantifying the characteristic dimension of the spatial
fluctuations. Finally, a random field is ergodic if all information on
joint PDFs are available from a single realization of the field.

With reference to the quantification of morphology of random
heterogeneous materials, variable $\theta$ simply denotes a
realization of random mesostructure drawn from the ensemble space
$\Theta$ of all admissible configurations. Of particular importance is
the characteristic function related to the spatial distribution of the
$i$-th phase:
\begin{equation}\label{eq1}
\chi\phs{i}(\mbf{x};\rl)
=\left\{\begin{array}{cl}
    1 & \mbox{if } \mbf{x}\in\Omega\phs{i}(\rl), \\
    0 & \mbox{otherwise},
    \end{array}\right.
\end{equation}
where $\Omega\phs{i}(\theta)$ is the domain occupied by the $i$-th
phase for realization $\rl$ and $i$ can take values $\{ s, m \}$,
where $s$ denotes the stone phase and $m$ refers to the mortar
phase. The characteristic functions of individual phases are not
independent, once e.g. the ``stone'' characteristic function is
provided, the complementary descriptor follows from
\begin{equation}
\chi\phs{m}(\mbf{x};\rl)
+
\chi\phs{s}(\mbf{x};\rl)
= 1.
\end{equation}
Therefore, we concentrate on the stone phase in the sequel.

When assuming the statistically uniform and ergodic media, the basic
spatial statistics is provided by
\begin{eqnarray}
\mean{\chi\phs{s}} = c\phs{s},
&&
\acorr{\chi\phs{s}}( \mbf{x} - \mbf{x}' )
=
\tppF{s}( \mbf{x} - \mbf{x}' ) - \left( c\phs{s} \right)^2,
\end{eqnarray}
where $c\phs{s}$ is the volume fraction of the relevant phase and
$\tppF{s}$ coincides with the two-point probability function, defined
for generic phases $i,j \in \{s,m\}$ as~\cite{Torq02}
\begin{equation}
\tppF{ij}( \mbf{x}, \mbf{x}' )
=
\EX{ \chi\phs{i}( \mbf{x}; \rl ) \chi\phs{j}( \mbf{x}'; \rl) },
\end{equation}
hence quantifying the probability of two points $\mbf{x}$ and
$\mbf{x}'$ being located in phases $i$ and $j$ (with $\tppF{i}$
abbreviating $\tppF{ii}$).

\begin{figure}[ht]
\centering
\includegraphics[width=\textwidth]{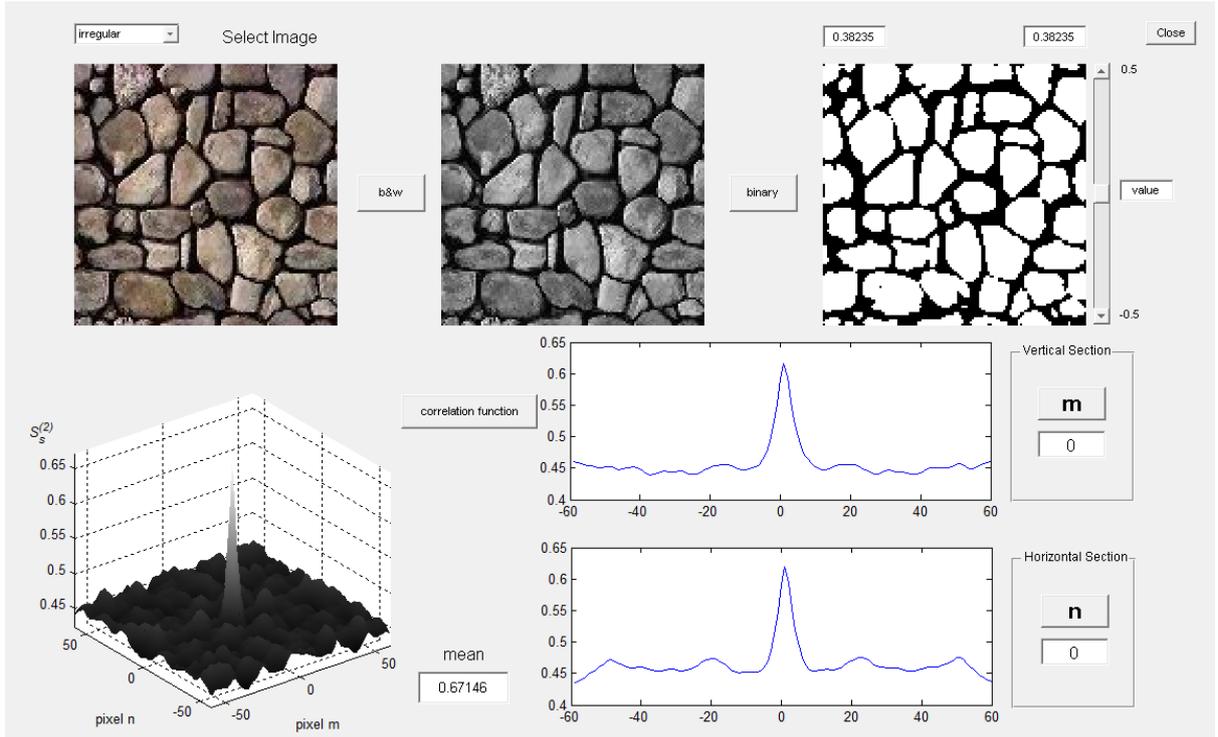}
\caption{Example of the program used to obtain correlation function
  for a chaotic masonry panel~(San Marco d'Alunzio, Italy).}
\label{figure1}
\end{figure}

The statistical descriptors of real mesostructures can be evaluated on
the basis of a digitized images of the sample, leading to the
discretization of the characteristic function in terms of an
$N_{x}\times{N_{y}}$ bitmap. Replacing the point coordinate $(x,y)$ by
the pixel $(i,j)$ located in the $i$-th row and the $j$-th column, the
characteristic function is defined by the discrete value
$\chi_s(i,j)$. The estimates of one-point and two-point correlation
functions, under the periodic boundary condition, follow from
relations~(see, e.g.,~\cite{GaZeSe06})
\begin{eqnarray}
c\phs{s}
& \approx &
\frac{1}{N_{x}N_{y}}\sum_{i=1}^{N_{x}}\sum_{j=1}^{N_{y}}\chi\phs{s}(i,j),
\label{eq:smp_oppf} \\
\tppF{s}(m,n)
& \approx &
\frac{1}{N_{x}N_{y}}
\sum_{i=1}^{N_{x}}\sum_{j=1}^{N_{y}}
\chi\phs{s}(i,j) \chi\phs{s}(1+(i+m)\%N_{x},1+(j+n)\%N_{y}),
\label{eq:smp_tppf}
\end{eqnarray}
where $m$ and $n$ are distances between two generic points measured in
pixels and $a\%b$ denotes $a$ modulo $b$. Note that the
sums~\eqref{eq:smp_oppf} and~\eqref{eq:smp_tppf} can be efficiently
evaluated using the Fast Fourier transform techniques; see
e.g.~\cite{Torq02}. To automate the acquisition of these functions, a
software working in {\bf MATLAB} was
implemented~\cite{Falsone:2007:SRMP}, covering all the basic steps of
mesostructure quantification with the data provided in the form of a
color image, see~\Fref{figure1} for an illustration of the procedure.

Apart from providing basic spatial statistics of random mesostructure,
the phase characteristic functions allow us to directly express the
matrix-valued field of material properties in the form
\begin{equation}\label{eq2}
\mbf{C}(\mbf{x};\rl)
=
\chi\phs{s}(\mbf{x};\rl)\mbf{C}\phs{s}
+
\chi\phs{m}(\mbf{x};\rl)\mbf{C}\phs{m}
=
\mbf{C}\phs{m} + \chi\phs{s}(\mbf{x};\rl) ( \mbf{C}\phs{s} - \mbf{C}\phs{m} ),
\end{equation}
where $\mbf{C}\phs{s}$ and $\mbf{C}\phs{m}$ are the deterministic
material stiffness matrices of the two constituents. The mean and the
covariance functions then follow from Equations~\eqref{eq:mean}
and~\eqref{eq:corr}:
\begin{eqnarray}
\mean{C_{ij}}( \mbf{x} )
& = &
C_{ij}\phs{m} + c\phs{s} ( C_{ij}\phs{s} - C_{ij}\phs{m} ),
\\
\corr{C_{ij}}{C_{kl}}( \mbf{x} - \mbf{x}' )
& = &
\acorr{\chi\phs{s}}( \mbf{x} - \mbf{x}' )
\left( C_{ij}\phs{s}-C_{ij}\phs{m} \right)
\left( C_{kl}\phs{s}-C_{kl}\phs{m} \right).
\end{eqnarray}
A similar representation is available when phase properties become
random variables, see~\cite{Falsone:2007:SRMP} for additional
discussion.

\section{Improved perturbation method}\label{Perturbative_approach}
Consider a mechanical system with the randomness in material
properties specified in terms of the random field
$\chi\phs{s}(\mbf{x};\rl)$. In the context of finite element analysis
of static problems, the discretized form of equilibrium equations
reads~\cite{Bittnar:1996:NMM}
\begin{equation}\label{eq17}
\mbf{K}\h(\chi\phs{s}(\mbf{x};\rl))
\mbf{u}\h(\rl)
=
\mbf{F}\h,
\end{equation}
where $h$ is a characteristic element size, the force vector
$\mbf{F}\h$ is assumed to be deterministic and the global stiffness
matrix $\mbf{K}\h$ is stochastic due to uncertainty in the material
properties, which makes the nodal displacement vector $\mbf{u}\h$
non-deterministic as well.

To handle the random field in \Eref{eq17} computationally, an
appropriate discretization technique in the stochastic variable has to
be used. In this work, the widely used mid-point method is employed to
represent the random field consistently with the underlying finite
element mesh. Therefore, two different considerations control the size
of an element~$h$, cf.~\cite{Babuska:2005:SEB}. The first one
comprises the usual ``deterministic'' criteria, where the mesh size is
governed by expected stress gradients and geometry. The additional
requirement is linked with the correlation length
$\lambda_{\chi\phs{s}}$; the distance between two adjacent random
variables has to be short enough to capture the essential features of
the random field. The general recommendation is to choose $2h \approx
\lambda_{\chi\phs{s}}$ to describe the stochastic field with
sufficient accuracy~\cite{Matthies:1997:UPN}.

In the current implementation, the value at the element center is used
to characterize the stochastic field, thus yielding a representation
in the form of a vector of random variables
\begin{equation}\label{eq18}
\mbf{\chi}\phs{s}\h(\theta)
=
\left[
 \begin{array}{cccc}
  \chi\phs{s}_{h,1}(\theta) &
  \chi\phs{s}_{h,2}(\theta) &
  \ldots &
  \chi\phs{s}_{h,N_e}(\theta)
 \end{array}
\right]\trn,
\end{equation}
with $\chi\phs{s}_{h,e}(\theta)$ being the value of
$\chi\phs{s}(\mbf{x};\rl)$ at the $e$-th element centroid and $N_e$
denoting the number of elements. The element stiffness matrix is
calculated from the standard finite element methodology and is
expressed as~\cite{Bittnar:1996:NMM}
\begin{equation}\label{eq19}
\mbf{K}_{h,e}\left( \chi\phs{s}_{h,e}(\theta) \right)
=
\int_{\Omega_e}
\mbf{B}_{h,e}\trn(\mbf{x})
\mbf{C}_e\left( \chi\phs{s}_{h,e}(\theta) \right)
\mbf{B}_{h,e}(\mbf{x})
\de \mbf{x},
\end{equation}
where $\mbf{B}_{h,e}$ is the deterministic displacement-to-strain matrix
related to the $e$-th element and the material stiffness matrix
$\mbf{C}_e$ follows from~\Eref{eq2}. After the assembly procedure, the
global form of equilibrium equations becomes
\begin{equation}\label{eq20}
\mbf{K}\h( \mbf{\chi}\phs{s}\h(\rl) )
\mbf{u}\h( \rl )
=
\mbf{F}\h.
\end{equation}

Among the various perturbative \SFE~approaches proposed in literature,
an improved perturbation technique proposed by Elishakoff et
al.~\cite{Elish95} is employed in this work. When compared to the
traditional first-order expansion schemes, the added value of the
adopted method is that the mean value of the response variables
depends on the covariance information on uncertain input parameters,
thereby optimally utilizing the available second-order statistics, see
also~\cite[Chapter~5]{Lombardo:2008:RFM} for further
discussion. Following this approach, the mean of the response vector
$\mbf{u}\h$ is given by
\begin{equation}\label{eq21}
\meanv{\mbf{u}\h}
=
\mbf{A}\h^{-1}
\mbf{F}\h,
\end{equation}
where
\begin{equation}\label{eq22}
\mbf{A}\h
=
\mbf{K}_{h,0}
-\sum_{i=1}^{N_e}\sum_{j=1}^{N_e}
\corr{\chi\phs{s}_i}{\chi\phs{s}_j}
\mbf{K}_{h,i}'
\left( \mbf{K_{h,0}}^{-1} \right)
\mbf{K}_{h,j}',
\end{equation}
with the stiffness matrix sensitivities provided by
\begin{eqnarray}
\mbf{K}_{h,0}
 =
\mbf{K}\h(\mbf{\chi}\phs{s}\h)
\Bigr|_{\mbf{\chi}\phs{s}\h=\EX{\mbf{\chi}\phs{s}\h}}, &&
\mbf{K}'_{h,i} = \frac{\partial
\mbf{K}\h(\mbf{\chi}\phs{s}\h)}{\partial \chi\phs{s}_{h,i}}
\Bigr|_{\mbf{\chi}\phs{s}\h=\EX{\mbf{\chi}\phs{s}\h}} .
\end{eqnarray}
In addition, the autocovariance matrix of displacements follows from
\begin{equation}\label{eq23}
\acorrm{\mbf{u}\h}
=
\mbf{K}_{0,h}^{-1}
\mbf{C}\h
\mbf{K_{0,h}}^{-1},
\end{equation}
where
\begin{equation}\label{eq24}
\mbf{C}\h
=
\sum_{i=1}^{N_e}
\sum_{j=1}^{N_e}
\corr{\chi\phs{s}_i}{\chi\phs{s}_j}
\mbf{K}_{h,i}'
\meanv{\mbf{u}\h}
\left( \meanv{\mbf{u}\h}\right)\trn
\mbf{K}_j',
\end{equation}
see~\cite{Elish95} for additional details.

\section{Karhunen-Lo\`{e}ve expansion}\label{KL-expansion}

The application of Karhunen-Lo\`{e}ve expansion~(\KLE) to stochastic
boundary value problems has been pioneered by Ghanem and his
co-workers~\cite{GhaSpa91b,GhaSpa91a} and provides an alternative way
to random field generation. The \KLE~can be seen as a special case of
the orthogonal series expansion where the orthogonal functions are
chosen as the eigenfunctions of a Fredholm integral equation of the
second kind with autocovariance as kernel~\cite{ZhaHel94}.

With reference to the mesostructure-based random fields considered in
the current work, we start from the \KLE~of the characteristic
function $\chi\phs{s}(\mbf{x};\theta)$ in the form
\begin{equation}\label{eq25}
\chi\phs{s}(\mbf{x};\theta)
=
\mean{\chi\phs{s}}(\mbf{x})
+
\sum_{i=1}^{\infty}
\sqrt{\lambda_i}\xi_i(\theta)f_i(\mbf{x}),
\end{equation}
where $\lambda_i$ and $f_i(\mbf{x})$ are the eigenvalues (decreasing
in magnitude) and eigenfunctions of the autocovariance
$\acorr{\chi\phs{s}}(\mbf{x},\mbf{x}')$, $\{\xi_i(\theta)\}$ is a set
of random variables~\cite{SudDer00}. Note that the use of \KLE~is
limited to the representation of \emph{input} random fields as the
covariance structure needs to be specified \emph{a priori}. Since the
kernel $\acorr{\chi\phs{s}}(\mbf{x},\mbf{x}')$ is bounded, symmetric
and non-negative, it has all eigenfunctions mutually orthogonal and
forming a complete set spanning the function space to which
$\chi\phs{s}(\mbf{x};\theta)$ belongs. Therefore, the autocovariance
function can be decomposed into
\begin{equation}\label{eq26}
\acorr{\chi\phs{s}}(\mbf{x},\mbf{x}')
=
\sum_{i=1}^{\infty}\lambda_i
f_i(\mbf{x})f_i(\mbf{x}'),
\end{equation}
with eigenfunctions $f_i(\mbf{x})$ and eigenvalues $\lambda_i$ found
as the solutions of the homogeneous Fredholm integral equation of the
second kind
\begin{equation}\label{eq28}
\int_\Omega
\acorr{\chi\phs{s}}(\mbf{x},\mbf{x}')
f_i(\mbf{x}')
\de \mbf{x}'
=
\lambda_i f_i(\mbf{x}).
\end{equation}
The parameter $\xi_i(\theta)$ in \Eref{eq25} corresponds to an
uncorrelated standardized random variable expressed as
\begin{equation}\label{eq29}
\xi_i(\theta)
=
\frac{1}{\sqrt{\lambda_i}}
\int_\Omega
\big[
 \chi\phs{s}(\mbf{x};\theta)-\mean{\chi\phs{s}}(\mbf{x})
\big]f_i(\mbf{x})
\de \mbf{x}.
\end{equation}
The most important aspect of the representation~\eqref{eq25} is that
the fluctuations are decomposed into a set of deterministic functions
in the spatial variables separately multiplying purely random
coefficients.

In practical implementations, the series~\eqref{eq25} and~\eqref{eq26}
are truncated after $M$ terms, yielding the approximations
\begin{eqnarray}
\chi\phs{s}(\mbf{x};\theta)
& \approx &
\mean{\chi\phs{s}}(\mbf{x})
+
\sum_{i=1}^{M}
\sqrt{\lambda_i}\xi_i(\theta)f_i(\mbf{x}),
\label{eq31}
\\
\acorr{\chi\phs{s}}(\mbf{x},\mbf{x}')
& \approx &
\sum_{i=1}^{M}
\lambda_i f_i(\mbf{x})f_i(\mbf{x}').
\label{eq32}
\end{eqnarray}
Such spatial semi-discretization is optimal in the sense that the mean
square error resulting from a truncation after the $M$-th term is
minimized~\cite{GhaSpa91a}.

The efficiency of \KLE~for simulating random fields crucially hinges
on accurate eigenvalues and eigenfunctions of the covariance
kernel. In this paper, the Galerkin method with orthogonal polynomial
basis functions is employed to solve the Fredholm
equation~\eqref{eq28} for general domains and autocovariance
functions, see \cite[Chapter~7]{Lombardo:2008:RFM} for implementation
details. In addition, a careful convergence study of truncated
\KLE~presented in~\cite{Huang01} has demonstrated, for specific
classes of stochastic fields, the dependence of the optimal value of
$M$ on the ratio of the characteristic domain length $L$ to the
correlation parameter $\lambda_{\chi\phs{s}}$. For weakly correlated processes
($\lambda_{\chi\phs{s}}/L \ll 1$), the higher order eigenvalues cannot be
neglected without having a serious impact on the accuracy of the
simulation.

\begin{figure}[ht]
\centering
\includegraphics[height=65mm]{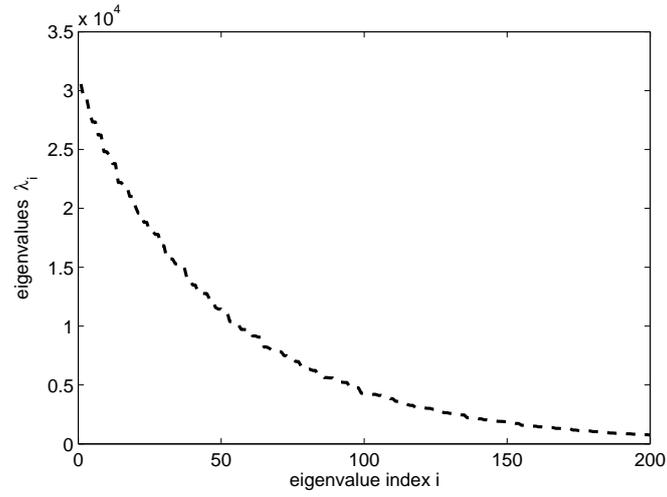}
\caption{Distribution of eigenvalues for \KLE~related to chaotic
  masonry sample.}
\label{figure2}
\end{figure}

\begin{figure}[ht]
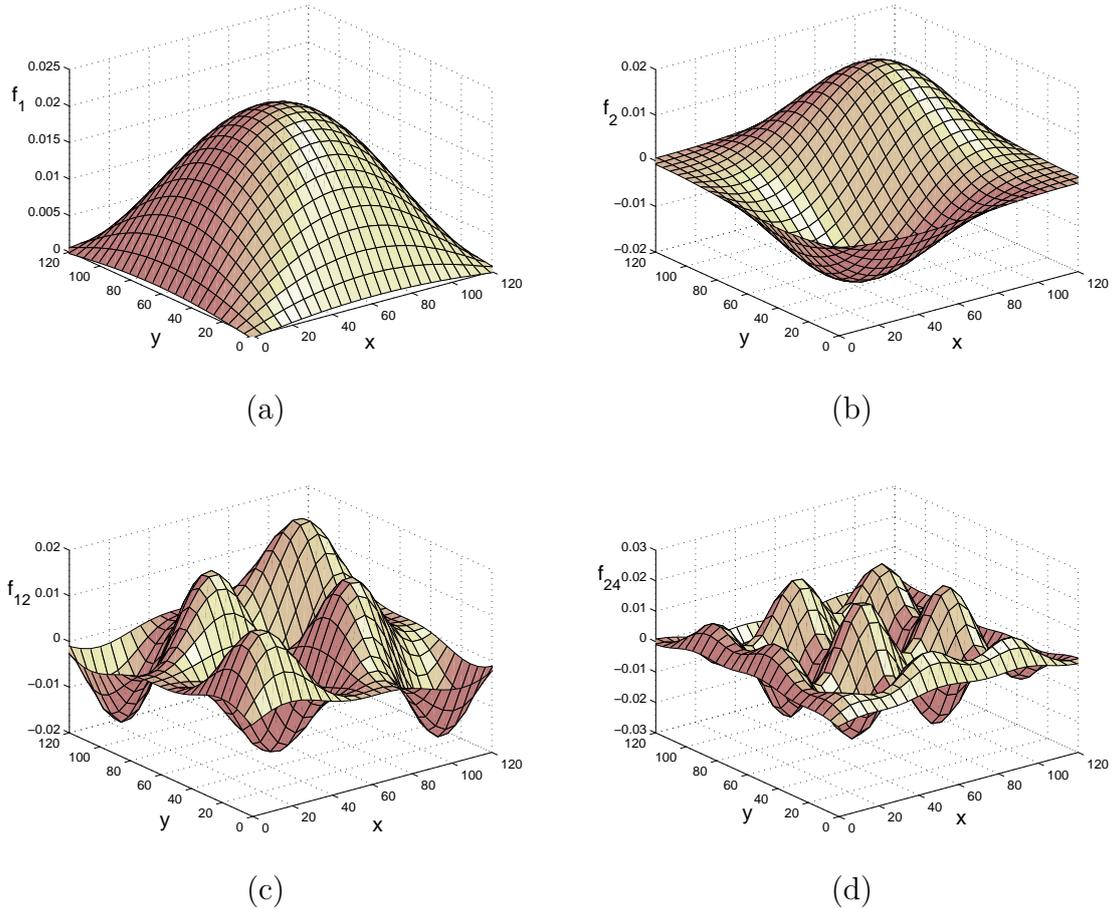

\centering
\subfigure[\label{figure3a}]%
 {\includegraphics[width=.45\textwidth]{\figName{figure3a}}}
\quad
\subfigure[\label{figure3b}]%
 {\includegraphics[width=.45\textwidth]{\figName{figure3b}}}
\\
\subfigure[\label{figure3c}]%
 {\includegraphics[width=.45\textwidth]{\figName{figure3c}}}
\quad
\subfigure[\label{figure3d}]%
 {\includegraphics[width=.45\textwidth]{\figName{figure3d}}}
\caption{Examples of eigenfunctions $f_i$; (a)~$i=1$, (b)~$i=2$,
  (c)~$i=12$, (d)~$i=24$.}
\label{figure3}
\end{figure}

Such behavior is illustrated by means of~\Fref{figure2}, showing the
decay of eigenvalues of \KLE~with the covariance kernel determined for
the masonry sample presented in~\Sref{Microstructural_statistics}. In
addition, several associated eigenfunctions are collected
in~\Fref{figure3}. Clearly, the random field under consideration is
weakly correlated as $\lambda_{\chi\phs{s}}/L \approx 10/120$, see \Fref{figure1},
and a large number of terms~($M \approx 200$) is needed to capture
fine features of the covariance, cf.~\Fref{figure3d}.

With a \KLE~of the spatial autocovariance function at hand, the
individual realizations of the heterogeneous body can be efficiently
generated once an appropriate model for the random field
$\chi\phs{s}(\mbf{x};\rl)$ is adopted. In the current study, we assume
that the random field is Gaussian, for which the coefficients
$\xi_i(\theta)$ in \Eref{eq29} become independent standard Gaussian
variables of zero mean and unit variance. It should be emphasized that
the assumption of Gaussian form of characteristic function is somehow
questionable for binary heterogeneous media, where the log-normal or
beta probability densities appear to be more appropriate to reflect
the intrinsic discreteness of the random field. The Gaussian
assumption is adopted here mainly due to simplicity of the resulting
simulation algorithm based on well-established routines, see also
related works~\cite{Charmpis:2007:NLM,Xu:2007:MSFEM,Xu:2005:SCM} for
further discussion.

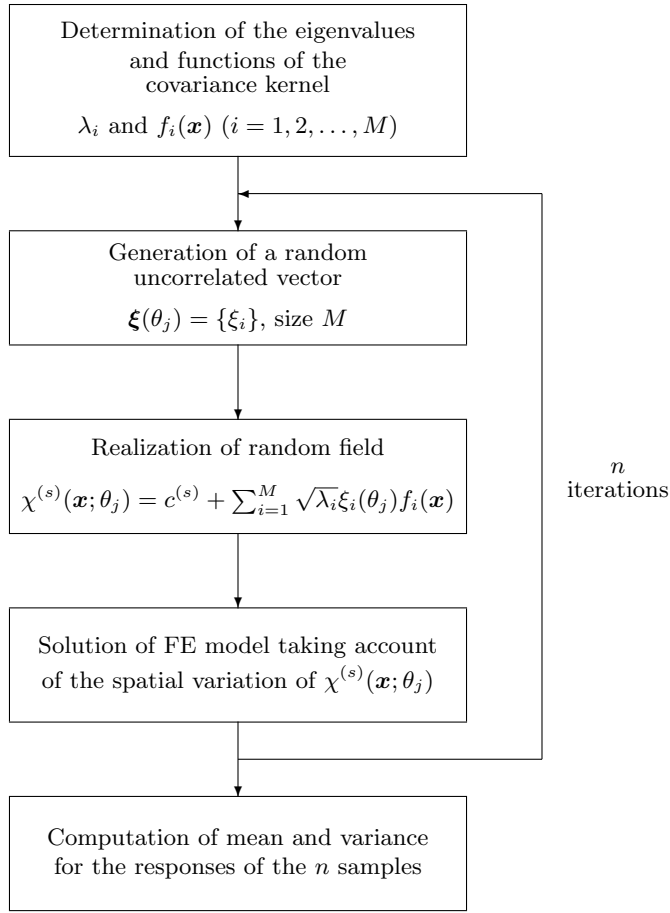
\begin{figure}[ht]
\centering
\scriptsize \setlength{\unitlength}{1mm}
\begin{picture}(140,130)(-70,-110)
\put(-30,0){\framebox(60,20)[c]{\shortstack[c]{ Determination of
the eigenvalues\\ and functions of the\\ covariance kernel\\[2mm]
$\lambda_i$ and $f_i(\mbf{x})$~($i=1,2,\dots,M$) }}}
\put(0,0){\vector(0,-10){10}}
 \put(40,-5){\vector(-1,0){40}}
 \put(40,-5){\line(0,-1){75}}
 \put(0,-80){\line(1,0){40}}
\put(-30,-25){\framebox(60,15)[c]{\shortstack[c]{Generation of a
random\\ uncorrelated vector\\[2mm]
$\mbf{\xi}(\theta_j)=\{\xi_i\}$, size $M$}}}
\put(0,-25){\vector(0,-10){10}}
\put(-30,-50){\framebox(60,15)[c]{\shortstack[c]{Realization of
random
field\\[3mm]
$\chi\phs{s}(\mbf{x};\theta_j)=c\phs{s}+\sum_{i=1}^M
\sqrt{\lambda_i}\xi_i(\theta_j)f_i(\mbf{x})$}}}
\put(0,-50){\vector(0,-10){10}}
\put(-30,-75){\framebox(60,15)[c]{\shortstack[c]{Solution of FE
model taking account\\ of the spatial variation of
$\chi\phs{s}(\mbf{x};\theta_j)$}}} \put(0,-75){\vector(0,-10){10}}
\put(-30,-100){\framebox(60,15)[c]{\shortstack[c]{Computation of
mean and variance\\ for the responses of the $n$ samples }}}
\put(48,-45){\makebox(4.0000,4.0000)[c]{\shortstack[c]{$n$\\
iterations }}}
\end{picture}
\caption{Overview of the Monte-Carlo simulation coupled with \KLE.}
\label{figure4}
\end{figure}

The final step of the \KLE-based solver involves the determination of
the response statistics for a structure with material stiffness
determined from~\Eref{eq2}:
\begin{equation}
\mbf{C}( \mbf{x}; \rl )
\approx
\mbf{C}\phs{m} + 
\left(
c\phs{s}
+
\sum_{i=1}^{M}
\sqrt{\lambda_i}\xi_i(\theta)f_i(\mbf{x})
\right)
( \mbf{C}\phs{s} - \mbf{C}\phs{m} ).
\end{equation}
The frequently adopted framework of spectral
\SFE~\cite{GhaSpa91a,Matthies:2005:GMLN}, where the response variable
is discretized using the polynomial chaos expansion in the stochastic
coordinate, is not applicable in the current case as the high number
of \KLE~terms results in unmanageable number of polynomial chaos
components. Therefore, similarly to
e.g.~\cite{Kowalsky:2007:RFM,Schenk:2003:BAC}, a direct Monte-Carlo
approach is adopted in the present study, leading to a simulation
procedure summarized in \Fref{figure4}. Note that in the FE analysis,
the characteristic element size $h$ has to verify $2h \leq
\lambda_{\phs{s}}$ again to correctly reproduce the autocovariance
function.

Once the sampling phase is completed, the unbiased mean and
covariance of displacement vectors are provided by
\begin{eqnarray}
\meanv{\mbf{u}\h}
& = &
\frac{1}{n}
\sum_{j=1}^{n}\mbf{u}\h(\theta_j),
\label{eq35} \\
\acorrm{\mbf{u}\h}
& = &
\frac{1}{n-1}
\sum_{j=1}^n
\left[
\mbf{u}\h(\theta_j)
\left( \mbf{u}\h(\theta_j) \right)\trn
-
n \meanv{\mbf{u}\h}
\left( \meanv{\mbf{u}\h} \right)\trn
\right]
\label{eq36},
\end{eqnarray}
where, in accord with \Fref{figure4}, $n$ denotes the number of
simulations and $\rl_j$ is used to denote the $j$-th deterministic
realization.

\section{Hashin-Shtrikman variational approach}\label{Hashin-Shtrikman}
%
The last approach investigated here builds on the classical
Hashin-Shtrikman variational principles for the heterogeneous
media~\cite{Hashin:1962a}, extended to the stochastic setting by
Willis~\cite{Willis:1977:BSC}. The basic idea of the method is the
introduction of a reference homogeneous body with stiffness tensor
$\mbf{C}_0$, employed in the analysis instead of an inhomogeneous
realization~$\mbf{C}(\mbf{x};\rl)$, recall \Eref{eq2}. The
heterogeneity of the material is compensated using the polarization
stress $\mbf{\tau}(\mbf{x};\theta)$, resulting from the stress
equivalence condition:
\begin{equation}\label{eq37}
\mbf{\sigma}(\mbf{x};\theta)
=
\mbf{C}(\mbf{x};\theta)
\mbf{\varepsilon}(\mbf{x};\theta)
=
\mbf{C}_0
\mbf{\varepsilon}(\mbf{x};\theta)
+
\mbf{\tau}(\mbf{x};\theta),
\end{equation}
with $\mbf{\sigma}$ and $\mbf{\varepsilon}$ denoting the
configuration-dependent stress and strain fields. The additional
unknown follows from stationarity conditions
\begin{equation}\label{eq38}
\big( \mbf{u}(\mbf{x};\theta), \mbf{\tau}(\mbf{x};\theta) \big)
=
\arg \min_{\mbf{v}(\mbf{x})} \stat_{\mbf{\omega}(\mbf{x})}
\Pi_{\HS}\left( \mbf{v}(\mbf{x}), \mbf{\omega}(\mbf{x}); \rl \right)
\end{equation}
where ``$\arg \min F$'' denotes the minimizer and ``$\arg \stat F$''
the stationary point of a functional $F$, respectively, and
$\Pi_{\HS}$ stands for the Hashin-Shtrikman (\HS) energy functional
provided by
\begin{eqnarray}
\Pi_{\HS}\left( \mbf{v}(\mbf{x}), \mbf{\omega}(\mbf{x}); \rl \right)
& = &
\half
\int_\Omega
\mbf{\varepsilon}\trn(\mbf{v}(\mbf{x}))
\mbf{C}_0
\mbf{\varepsilon}(\mbf{v}(\mbf{x}))
\de \mbf{x}
-
\int_\Omega
\mbf{v}\trn( \mbf{x} ) \mbf{f}( \mbf{x} )
\de \mbf{x}
\nonumber \\
& - &
\int_{\Gamma_t}
\mbf{v}\trn( \mbf{x} ) \mbf{t}( \mbf{x} )
\de \mbf{x}
+
\int_\Omega
\mbf{\varepsilon}\trn(\mbf{v}(\mbf{x}))
\mbf{\omega}(\mbf{x})
\de \mbf{x}
\nonumber \\
& - &
\half
\int_\Omega
\mbf{\omega}\trn( \mbf{x} )
\big[\mbf{C}(\mbf{x};\theta)-\mbf{C}_0\big]^{-1}
\mbf{\omega}( \mbf{x} )
\de \mbf{x}.
\label{eq39}
\end{eqnarray}
In \Eref{eq38}, $\mbf{v}$ and are $\mbf{\omega}$ denote trial values
of displacement field and polarization stresses, while
$\mbf{f}(\mbf{x})$ are deterministic body forces and
$\mbf{t}(\mbf{x})$ boundary tractions acting on $\Gamma_t$,
respectively, cf.~\cite{LucWil05}. It can be shown that when the
optimization in~\eqref{eq38} is performed exactly, the stationary
value of functional~$\Pi_{\HS}$ coincides with the actual energy
stored in the system for realization~$\rl$. Moreover, it
holds~\cite{Hashin:1962a,Prochazka:2004:EHS}
\begin{equation}\label{eq:HS_ineq}
\Pi_{\HS} \left( \mbf{u}(\mbf{x};\rl), \mbf{\omega}(\mbf{x}); \rl \right)
\leq
\Pi_{\HS}\left( \mbf{u}(\mbf{x};\rl), \mbf{\tau}(\mbf{x};\rl); \rl \right)
\end{equation}
whenever $(\mbf{C}(\mbf{x};\theta)-\mbf{C}_0)$ is positive-definite;
when $\mbf{C}_0$ is chosen such that the difference becomes
negative-definite, the inequality is reversed.

The elementary statistics of displacements and polarizations
associated with probability density $\pme(\rl)$ follow directly from a
stochastic variant of \Eref{eq38}:
\begin{equation}\label{eq:stoch_var_princ}
\big( \meanv{\mbf{u}}, \meanv{\mbf{\tau}} \big)
=
\int_\Theta
\left(
\arg \min_{\mbf{v}(\mbf{x};\rl)} \stat_{\mbf{\omega}(\mbf{x};\rl)}
\Pi_{\HS}\left( \mbf{v}(\mbf{x};\rl), \mbf{\omega}(\mbf{x};\rl); \rl \right)
\right)
\de \pme(\rl).
\end{equation}
Following the approach of Willis~\cite{Willis:1977:BSC}, the previous
problem is solved approximately by considering the following ansatz
for displacements and polarizations:
\begin{eqnarray}
\mbf{u}(\mbf{x},\theta)
=
\mbf{u}_0(\mbf{x})+\mbf{u}_1(\mbf{x};\theta),
&&
\mbf{\tau}(\mbf{x};\theta)
\approx
\chi\phs{s}(\mbf{x};\theta)\mbf{\tau}\phs{s}(\mbf{x})
+
\chi\phs{m}(\mbf{x};\theta)\mbf{\tau}\phs{m}(\mbf{x}),
\end{eqnarray}
where $\mbf{u}_0$ is the deterministic displacement of the reference
body subject to distributed body forces and tractions, $\mbf{u}_1$
stores the configuration-dependent displacement due to the
polarization stress expressed using a non-local operator
$\mbf{\Gamma}_0$, cf.~\Eref{eq41} bellow,
\begin{equation}
\mbf{u}_1( \mbf{x}; \rl )
=
- \int_\Omega
\mbf{\Gamma}_0(\mbf{x},\mbf{x'}) 
\mbf{\tau}(\mbf{x}';\rl)
\de \mbf{x}',
\end{equation}
and $\mbf{\tau}\phs{i}$ denotes the deterministic value of the
polarization stress related to the $i$-th phase. 

In accord with the standard Galerkin procedure, an identical form is
adopted for the trial values of displacements and polarizations. Upon
exchanging the order of optimization, the optimality conditions of
problem~\eqref{eq:stoch_var_princ} reduce to
identity~\cite{LucWil05,LucWil06}
\begin{eqnarray}\label{eq40}
\int_\Omega
\mbf{\omega}\phs{i}\trn(\mbf{x})
c\phs{i}
\big[\mbf{C}\phs{i}-\mbf{C}_0\big]^{-1}
\mbf{\tau}\phs{i}(\mbf{x})
\de \mbf{x}
& + & \\
\sum_j
\int_\Omega
\mbf{\omega}\phs{i}\trn(\mbf{x})
\int_\Omega
\tppF{ij}( \mbf{x} - \mbf{x}' )
\mbf{\Gamma}_0(\mbf{x},\mbf{x'})
\mbf{\tau}\phs{j}(\mbf{x}')
\de \mbf{x}'
& = &
\int_\Omega
\mbf{\omega}\phs{i}(\mbf{x})\trn
c\phs{i}
\mbf{\varepsilon}(\mbf{u}_0(\mbf{x}))
\de \mbf{x},
\nonumber
\end{eqnarray}
to be satisfied for an arbitrary $\mbf{\omega}\phs{i}$ and $i,j \in
\{s,m\}$. 

Two levels of approximation are generally needed to fully discretize
the system~\eqref{eq40}. The first step involves discretizing the
$\mbf{\Gamma}_0$ operator together with the ``reference'' strain
distribution, which in the context of the adopted Finite Element
approximation become~\cite{LucWil05,LucWil06}
\begin{eqnarray}\label{eq41}
\mbf{\Gamma}_0(\mbf{x},\mbf{x}')
\approx
\mbf{B}\h\u(\mbf{x})
\mbf{K}_{h,0}^{-1}
\mbf{B}\h\u\trn(\mbf{x'}),
&
\mbf{\varepsilon}(\mbf{u}_0(\mbf{x}))
\approx
\mbf{B}\h\u(\mbf{x}) \mbf{u}_{0,h},
&
\mbf{u}_0(\mbf{x})
\approx
\mbf{N}\h\u(\mbf{x}) \mbf{u}_{0,h},
\end{eqnarray}
where, in analogy with \Sref{Perturbative_approach}, $\mbf{K}_{h,0}$
denotes the stiffness matrix of the reference structure, $\mbf{N}\h\u$
is the matrix of shape functions and $\mbf{B}\h = \mbf{\partial}
\mbf{N}\h\u$ is the displacement-to-strain matrix and $\mbf{u}_{0,h}$
stands for nodal displacement vector determined for the reference
problem~\cite{Bittnar:1996:NMM}. In the second step, phase
polarization fields are parameterized in the form, cf.~\cite{LucWil05}
\begin{equation}\label{eq42}
\mbf{\tau}\phs{i}(\mbf{x})
\approx
\mbf{N}\h\t(\mbf{x})
\mbf{d}\h\phs{i},
\end{equation}
where $\mbf{N}\t\h$ is the matrix of shape functions to approximate
the polarization stresses. It is worth mentioning that a detailed
one-dimensional study presented in~\cite{Sharif-Khodaei:2008:MBM}
demonstrated that, similarly to the remaining approaches, the
characteristic element size $2h \approx \lambda_{\chi\phs{s}}$ is
again necessary to achieve a sufficient accuracy of the obtained
statistics.

Employing the approximations~\eqref{eq41} and~\eqref{eq42},
the stationary conditions~\eqref{eq40} yield the system of linear
equations
\begin{equation}\label{eq:discr_system}
\mbf{K}\h\phs{i}\mbf{d}\phs{i}\h
+
\sum_j
\mbf{K}\h\phs{ij}
\mbf{d}\h\phs{j}
=
\mbf{R}\h\phs{i},
\end{equation}
with the individual terms provided by
\begin{eqnarray}\label{eq43}
\mbf{K}\h\phs{i}
& = &
\int_\Omega
\mbf{N}\t\h\trn(\mbf{x})
c\phs{i}
\big[\mbf{C}\phs{i}-\mbf{C}_0\big]^{-1}
\mbf{N}\t\h(\mbf{x})
\de \mbf{x},
\\
\mbf{K}\h\phs{ij}
&=&
\int_\Omega \int_\Omega
\mbf{N}\t\h\trn(\mbf{x})
\tppF{ij}(\mbf{x}-\mbf{x'})
\mbf{\Gamma}_{0,h}(\mbf{x},\mbf{x'})
\mbf{N}\t\h(\mbf{x}')
\de \mbf{x} \de \mbf{x}',
\\
\mbf{R}\phs{i}\h
&=&
\int_\Omega
\mbf{N}\t\h\trn(\mbf{x})
c\phs{i}
\mbf{B}\h\u(\mbf{x}) \mbf{u}_{0,h}
\de \mbf{x}.
\end{eqnarray}
Once the degrees of freedom related to the phase polarization stresses
are determined from system~\eqref{eq:discr_system}, the mean of
displacement value becomes~\cite{LucWil05}
\begin{equation}
\meanv{\mbf{u}\h}( \mbf{x} )
=
\mbf{N}^u\h( \mbf{x} )
\left( 
\mbf{u}_{0,h}
-
\mbf{K}_{0,h}^{-1}
\int_\Omega
\mbf{B}^u\h\trn( \mbf{x}' )
\meanv{\mbf{\tau}\h}( \mbf{x}' )
\de \mbf{x}'
\right),
\end{equation}
with
\begin{equation}
\meanv{\mbf{\tau}\h}( \mbf{x}' )
=
\mbf{N}^\tau\h(\mbf{x}')
\left(
c\phs{m} \mbf{d}\h\phs{m}
+
c\phs{s} \mbf{d}\h\phs{s}
\right).
\end{equation}
In addition to the mean response, the \HS~approach offers an
alternative way to establishing confidence-like bounds on the expected
displacements by varying the auxiliary stiffness $\mbf{C}_0$. In
particular, it follows from~\Eref{eq:HS_ineq} that selecting the
reference medium such that $\mbf{C}_0 = \min_i( \mbf{C}\phs{i} )$
yields an upper bound of the stored energy~(and therefore the upper
``energetic'' bounds of the displacements), whereas the choice
$\mbf{C}_0 = \max_i( \mbf{C}\phs{i} )$ results in a lower bound on
displacements. Finally, selecting $\mbf{C}_0$ such that the difference
$(\mbf{C}-\mbf{C}_0)$ becomes indefinite provides general variational
estimates of the basic statistics, cf.~\cite{Dvorak:1999:NEOP}.

\section{Numerical example}\label{Numerical_Example}
\begin{figure}[ht]
\centering
\includegraphics[height=65mm]{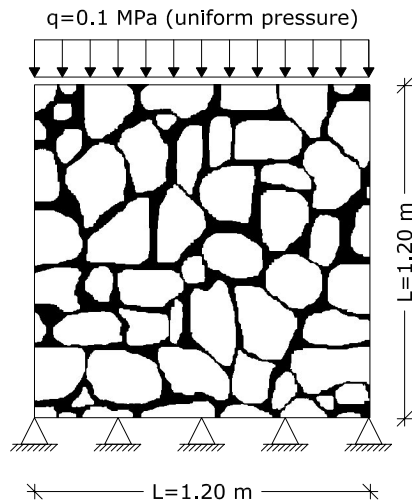}
\caption{Scheme of an illustrative example}
\label{figure5}
\end{figure}

In this Section, the essential features of the proposed numerical
methods are illustrated by studying elastic response of an irregular
masonry structure with dimensions shown in \Fref{figure5} and constant
thickness of $0.12$~m. The plane stress assumptions were adopted in
the analysis; the structure was subject to a uniform pressure applied
at the top edge and to the self-weight (a deterministic specific
gravity equal to $20$~kNm$^{-3}$ was assumed for simplicity). Material
constants of individual constituents were considered to be
deterministic, the concrete values of the Young moduli $E\phs{m} =
1,200$~MPa, $E\phs{s} = 12,500$~MPa and of the Poisson ratios
$\nu\phs{m}= 0.3$ and $\nu\phs{s}= 0.2$ were selected
following~\cite{Cluni:2004:HNP}. The geometrical uncertainty due to
irregular configuration of individual phases was quantified on the
basis of image analysis data presented in \Fref{figure1}.

The finite element model of the example problem was based on a regular
discretization of the domain using $24 \times 24$ square bilinear
elements with four integration points. Note that such a resolution
corresponds to the element edge approximately equal to a half of the
geometrical correlation length, which is fully consistent with general
rules discussed in
Sections~\ref{Perturbative_approach}--\ref{Hashin-Shtrikman}.

\begin{figure}[ht]
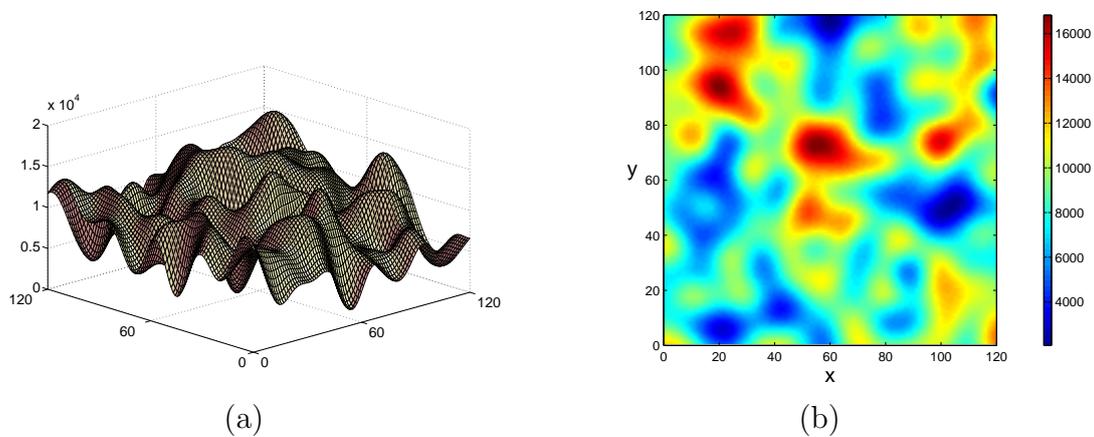

\centering
\subfigure[\label{figure6a}]%
 {\includegraphics[width=.45\textwidth]{\figName{figure6a}}}%
\quad
\subfigure[\label{figure6b}]%
 {\includegraphics[width=.45\textwidth]{\figName{figure6b}}}%
\caption{Realization of Gaussian random field of Young's modulus}
\label{figure6}
\end{figure}

The results presented for the \KLE-based solver were derived from
$n=1,000$ simulations. For simplicity, only the Young modulus $E$
considered in the form of a random field~(see \Fref{figure6} for an
illustration), whereas the Poisson ratio was set to a deterministic
value determined from the Voigt estimate of the homogenized stiffness
matrix
\begin{equation}\label{eq:voigt_estim}
\meanv{\mbf{C}}
=
c\phs{m} \mbf{C}\phs{m} 
+ 
c\phs{s} \mbf{C}\phs{s}.
\end{equation}
Finally, based on a systematic one-dimensional study of the
\HS~approach presented in~\cite{Sharif-Khodaei:2008:MBM}, the
element-wise constant discretization of the phase polarization
stresses with four-point integration scheme was adopted to ensure
sufficient resolution of the spatial statistics. Additional
implementation-related details can be found
in~\cite{Lombardo:2008:RFM}.

\begin{figure}[hbt]
\centering
\includegraphics*[height=70mm]{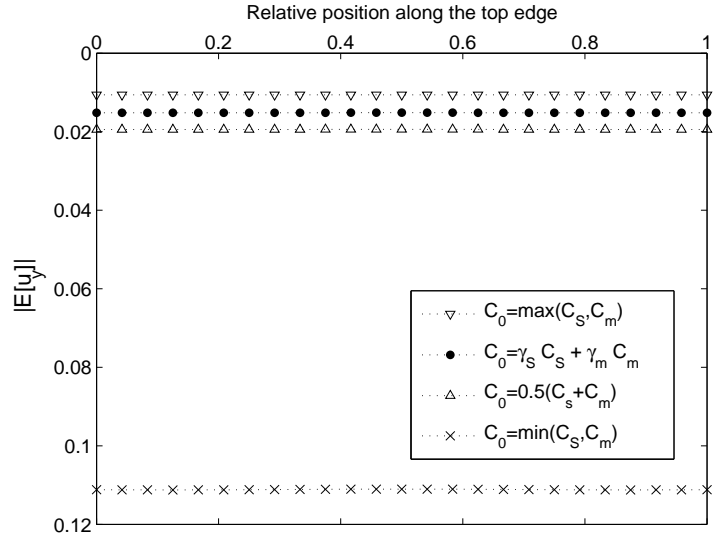}
\caption{Expected value of nodal displacements via \HS-based FE
  analysis}
\label{figure7}
\end{figure}

Before presenting the comparison of individual approaches, we
concentrate first on the effect of reference media on the \HS-based
predictions. To this end, the expected values of nodal displacements
are plotted in \Fref{figure7} for several representative choices of
$\mbf{C}_0$. In particular, owing to the dominant fraction of the
stiffer phase~($c\phs{s} \doteq 67\%$) in the considered
structure, the lower energetic bound can be expected to be
substantially closer to the ``true'' statistics than the corresponding
upper bound, which in the current case seems to be too inaccurate for
practical use. Additional estimates can be generated by the Voigt-type
choice~\eqref{eq:voigt_estim} or by setting the reference medium to
the arithmetic average of properties of individual constituents, the
value commonly adopted in the polarization-based numerical method due
to Moulinec and Suquet~\cite{Moulinec:1998:NMC}. As expected, the
response corresponding to such choices is comparable to the lower
bound and will be used in the sequel for the comparison with the
candidate approaches.

\begin{figure}[ht]
\centering
\includegraphics*[height=70mm]{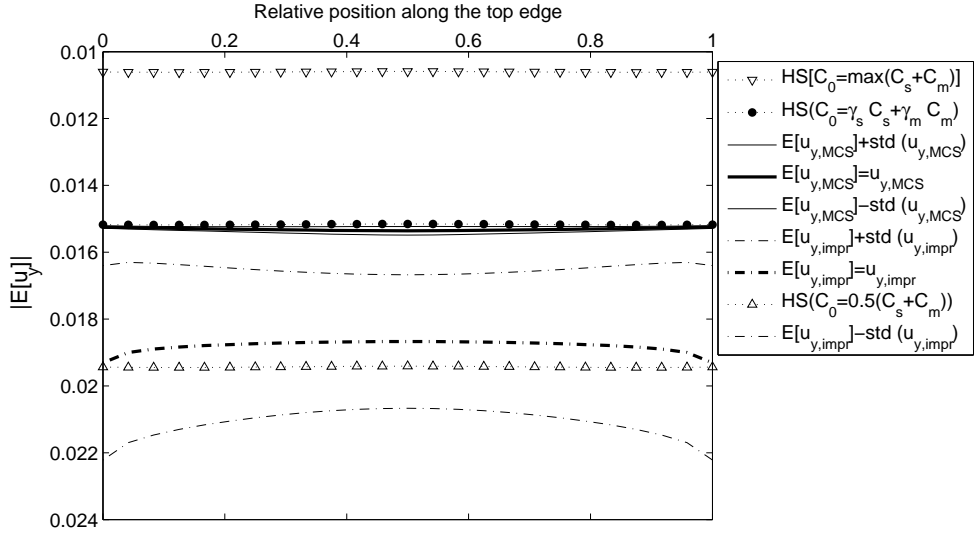}
\caption{Basic statistics of nodal displacements on the top of the panel}
\label{figure8}
\end{figure}

The basic statistics of nodal displacements, as predicted by different
methods, are mutually compared in \Fref{figure8}. In addition, we
present the confidence bounds in the form $\meanv{\mbf{u}\h}(\mbf{x})
\pm \mbf{\sigma}_{\mbf{u}\h}(\mbf{x})$, determined on the basis of the
second-order statistics for the improved perturbation
method~\eqref{eq24} or~\KLE~\eqref{eq36}. In general, it can be seen
that the perturbative method leads to a substantially wider confidence
interval when compared to the Monte-Carlo simulation approach, in
spite of a moderate number of simulations used by \KLE~solver to
estimate the overall statistics. For both methods, the confidence
intervals remain bounded from above by the corresponding
\HS~value. Moreover, for appropriate choices of the reference
stiffness matrix, the \HS~method recovers the predictions provided by
the alternative approaches. For the current setting, selecting
$\mbf{C}_0$ according to the rule of mixtures yields the displacement
values almost identical to that of the \KLE~solver, whereas the
response related to the arithmetic average well approximates the
improved perturbation result. These results provide just another
highlight of the importance of a proper choice of the reference media
in the \HS-based schemes, see
e.g.~\cite{Dvorak:1999:NEOP,Sharif-Khodaei:2008:MBM} for further
discussion.

The final comment concerns the computational complexity of individual
approaches. It can be stated that the cost of the improved
perturbation method and the \HS-based solver is roughly comparable,
whereas the \KLE~approach leads to an approximately three-fold
increase in the simulation time. Higher cost of the latter method can
be attributed to a large number of terms appearing in the
expansion~\eqref{eq31}; the computational cost, however, is
compensated by generality of the Monte-Carlo framework and can be
further reduced by parallelization of the problem.

\section{Conclusions and future work}\label{Conclusions}

In this contribution, the applicability of three distinct approaches
to mesostructure-based random field simulation of irregular historic
masonry was investigated. The numerical results obtained for a
finite-size elastic panel allow us to reach the following conclusions:
\begin{itemize}

\item The elements of quantification of random spatial statistics can
  be efficiently used to construct realistic first- and second-order
  statistics of stationary random fields.

\item The improved perturbation method utilizes the second-order
  statistics when determining the mean response of the system, which
  generally leads to narrower estimates when compared with the basic
  method, e.g.~\cite[Chapter~5]{Lombardo:2008:RFM}. In the current
  case, however, the uncertainty in the obtained statistics is higher
  than for the \KLE~algorithm, mainly due to a relatively high
  contrast of phase stiffnesses.

\item The Karhunen-Lo\`{e}ve series representation coupled with the
  Monte Carlo approach provides an interesting alternative to the
  perturbation-based method, even at increased computational
  cost. When applied to realistic structures, however, a large number
  of terms seems to be necessary to capture the available covariance
  information. Moreover, the validity of the Gaussian assumption needs
  to be critically assessed.

\item The Hashin-Shtrikman approach takes advantage of the specific
  form of the random field and therefore optimally utilizes the
  available information. The overall response is in this case,
  however, highly dependent on the choice of the reference medium, for
  which there is no general rule.

\end{itemize}

Even though the results of this pilot study have provided valuable
insights into the pros and cons of individual methods, they do not
allow for directly quantifying the accuracy of individual methods as
the reference solution is not available. Similarly to a recent
study~\cite{Sharif-Khodaei:2008:MBM}, such a comparison can be based
on a synthetic mesostructural model such the one proposed by Spence at
al.~\cite{Spence:2008:PMS}. This particular topic enjoys our current
interest and will be reported separately.

\section*{Acknowledgements}
JZ would like to express his sincere thanks to Oliver Allix, LMT
Cachan, for bringing his attention to the problem of \KLE~based on
spatial statistics. In addition, JZ and M\v{S} gratefully acknowledge
support of this research from the research project MSM~6840770003
(M\v{S}MT \v{C}R). The work of ML as a visiting research student at
the Department of Mechanics at Faculty of Civil Engineering of the
Czech Technical University in Prague was partially supported by a
grant for Doctoral Studies provided by the University of
Messina. Moreover, ML would like to express her gratitude to all the
academics and students who made her stay in Prague a very fruitful
period.


\begin{thebibliography}{10}

\bibitem{Anthoine:1995:DIPE} 
A.~Anthoine.  Derivation of the in-plane elastic characteristics of
masonry through homogenization theory.  {\em International Journal of
  Solids and Structures}, 32(2):137--163, 1995.

\bibitem{Babuska:2005:SEB}
I.~Babu\v{s}ka, R.~Tempone, and G.~E. Zouraris.  Solving elliptic
boundary value problems with uncertain coefficients by the finite
element method: the stochastic formulation.  {\em Computer Methods in
  Applied Mechanics and Engineering}, 194(12--16):1251--1294, 2005.

\bibitem{Bensoussan:1978:AAPS}
A.~Bensoussan, J.L. Lions, and G.~Papanicolaou.  {\em Asymptotic
  analysis for periodic structures}.  Studies in Mathematics and its
Applications. North-Holland, Amsterdam, 1978.

\bibitem{Bittnar:1996:NMM}
Z.~Bittnar and J.~\v{S}ejnoha.  {\em Numerical methods in structural
  mechanics}.  ASCE Press and Thomas Telford, Ltd., New York and
London, 1996.

\bibitem{Buryachenko:2007:MHM}
V.~Buryachenko.  {\em Micromechanics of heterogeneous materials}.
Springer Verlag, New York, NY, USA, 2007.

\bibitem{Charmpis:2007:NLM}
D.~C. Charmpis, G.~I. Schueller, and M.~F. Pellissetti.  The need for
linking micromechanics of materials with stochastic finite elements: A
challenge for materials science.  {\em Computational Materials
  Science}, 41(1):27--37, 2007.

\bibitem{Cluni:2004:HNP}
F.~Cluni and V.~Gusella.
Homogenization of non-periodic masonry structures.
{\em International Journal of Solids and Structures},
  41(7):1911--1923, 2004.

\bibitem{Dvorak:1999:NEOP}
G.~J. Dvorak and M.~V. Srinivas.
New estimates of overall properties of heterogeneous solids.
{\em Journal of the Mechanics and Physics of Solids}, 47(4):899--920,
  1999.

\bibitem{Elish95}
I.~Elishakoff, Y.~J. Ren, and M.~Shinozuka.
Improved finite element method for stochastic problems.
{\em Chaos, Solitons \& Fractals}, 5(5):833--846, 1995.

\bibitem{Falsone:2007:SRMP}
G.~Falsone and M.~Lombardo.
Stochastic representation of the mechanical properties of irregular
  masonry structures.
{\em International Journal of Solids and Structures},
  44(25--26):8600--8612, 2007.

\bibitem{GaZeSe06}
J.~Gajdo\v{s}\'{\i}k, J.~Zeman, and M.~\v{S}ejnoha.
Qualitative analysis of fiber composite microstructure: {I}nfluence
  of boundary conditions.
{\em Probabilistic Engineering Mechanics}, 21(4):317--329, 2006.

\bibitem{GhaSpa91b}
R.~Ghanem and P.~D. Spanos.
Spectral stochaststic finite element formulation for reability
  analysis.
{\em Journal of Engineering Mechanics ASCE}, 117(10):2351--2372,
  1991.

\bibitem{GhaSpa91a}
R.~Ghanem and P.~D. Spanos.
{\em Stochastic finite elements: {A} spectral approach}.
Dover Publications, Mineola, New York, second revised edition, 2003.

\bibitem{Gusella:2006:RFH}
V.~Gusella and F.~Cluni.
Random field and homogenization for masonry with nonperiodic
  microstructure.
{\em Journal of Mechanics of Materials and Structures},
  1(2):357--386, 2006.

\bibitem{Hashin:1962a}
Z.~Hashin and S.~Shtrikman.
On some variational principles in anisotropic and nonhomogeneous
  elasticity.
{\em Journal of the Mechanics and Physics of Solids}, 10:335--342,
  1962.

\bibitem{Huang01}
S.~P. Huang, S.~T. Quek, and K.~K. Phoon.
Convergence study of truncated {K}arhunen-{L}o\`{e}ve expansion for
  simulation of stochastic process.
{\em International Journal for Numerical Methods in Engineering},
  52(9):1029--1043, 2001.

\bibitem{Huet:1990:AVC}
C.~Huet.
Application of variational concepts to size effects in elastic
  heterogeneous bodies.
{\em Journal of the Mechanics and Physics of Solids}, 38(6):813--841,
  1990.

\bibitem{Jardak:2004:SSH}
M.~Jardak and R.~G. Ghanem.
Spectral stochastic homogenization of divergence-type {PDE}s.
{\em Computer Methods in Applied Mechanics and Engineering},
  193(6-8):429--447, 2004.

\bibitem{Jikov:1994:HDE}
V.~V. Jikov, S.~M. Kozlov, and O.~A. Oleinik.
{\em Homogenization of Differential Operators and Integral
  Functionals}.
Springer-Verlag, 1994.

\bibitem{Kaminski:1996:HER}
M.~Kami\'{n}ski.
Homogenization in elastic random media.
{\em Computer Assisted Mechanics and Engineering Sciences},
  3(1):9--21, 1996.

\bibitem{Kaminski:2004:CMCM}
M.~Kami\'{n}ski.
{\em Computational Mechanics of Composite Materials: {S}ensitivity,
  Randomness and Multiscale Behaviour}.
Springer Verlag, London, 2004.

\bibitem{Kaminski:2000:PBSFEM}
M.~Kami\'{n}ski and M.~Kleiber.
Perturbation based stochastic finite element method for
  homogenization of two-phase elastic composites.
{\em Computers \& Structures}, 78(6):811--826, 2000.

\bibitem{Kowalsky:2007:RFM}
U.~Kowalsky, T.~Zumendorf, and D.~Dinkler.
Random fluctuations of material behaviour in {FE}-damage analysis.
{\em Computational Materials Science}, 39(1):8--16, 2007.

\bibitem{Lombardo:2008:RFM}
M.~Lombardo.
{\em Random field models and stochastic analysis of irregular masonry
  structures}.
PhD thesis, Univesit\`{a} degli Studi di Messina, Facolt\`{a} di
  Ingegneira, Dipartmento di Ingegneira Civile, Messina, Italy, 2008.

\bibitem{Lourenco:2007:AMS}
P.B. Lourenco, G.~Milan, A.~Tralli, and Zucchini A.
Analysis of masonry structures: {R}eview of and recent trends in
  homogenization techniques.
{\em Canadian Journal of Civil Engineering}, 34(11):1443--1457, 2007.

\bibitem{LucWil05}
R.~Luciano and J.~R. Willis.
{FE} analysis of stress and strain fields in finite random composite
  bodies.
{\em Journal of the Mechanics and Physics of Solids},
  53(7):1505--1522, 2005.

\bibitem{LucWil06}
R.~Luciano and J.R. Willis.
Hashin-{S}htrikman based {FE} analysis of the elastic behaviour of
  finite random composite bodies.
{\em International Journal of Fracture}, 137(1--4):261--273, 2006.

\bibitem{Massart:2007:EMS}
T.J. Massart, R.H.J. Peerlings, and M.G.D. Geers.
An enhanced multi-scale approach for masonry wall computations with
  localization of damage.
{\em International Journal for Numerical Methods in Engineering},
  69(5):1022--1059, 2007.

\bibitem{Matthies:1997:UPN}
H.G. Matthies, C.E. Brenner, C.~G. Bucher, and C.~G. Soares.
Uncertainties in probabilistic numerical analysis of structures and
  solids: {S}tochastic finite elements.
{\em Structural Safety}, 19(3):283--336, 1997.

\bibitem{Matthies:2005:GMLN}
H.G. Matthies and A.~Keese.
Galerkin methods for linear and nonlinear elliptic stochastic partial
  differential equations.
{\em Computer Methods in Applied Mechanics and Engineering},
  194(12--16):1295--1331, 2005.

\bibitem{Moulinec:1998:NMC}
H.~Moulinec and P.~Suquet.
A numerical method for computing the overall response of nonlinear
  composites with complex microstructure.
{\em Computer Methods in Applied Mechanics and Engineering},
  157(1--2):69--94, 1998.

\bibitem{Povirk:1995:IMI}
G.~L. Povirk.
Incorporation of microstructural information into models of two-phase
  materials.
{\em Acta Metallurgica et Materialia}, 43(8):3199--3206, 1995.

\bibitem{Prochazka:2004:EHS}
P.~Proch\'{a}zka and J.~\v{S}ejnoha.
Extended {H}ashin-{S}htrikman variational principles.
{\em Applications of Mathematics}, 49(4):357--372, 2004.

\bibitem{Sab:1992:HSRM}
K.~Sab.
On the homogenization and the simulation of random materials.
{\em European Journal of Mechanics A-Solids}, 11(5):585--607, 1992.

\bibitem{Sakata:2008:3DSA}
S.~Sakata, F.~Ashida, T.~Kojima, and M.~Zako.
Three-dimensional stochastic analysis using a perturbation-based
  homogenization method for elastic properties of composite material
  considering microscopic uncertainty.
{\em International Journal of Solids and Structures},
  45(3--4):894--907, 2008.

\bibitem{Sakata:2008:KBA}
S.~Sakata, F.~Ashida, and M.~Zako.
Kriging-based approximate stochastic homogenization analysis for
  composite materials.
{\em Computer Methods in Applied Mechanics and Engineering},
  197(21--24):1953--1964, 2008.

\bibitem{Schenk:2003:BAC}
C.A. Schenk and G.I. Schu\"{e}ller.
Buckling analysis of cylindrical shells with random geometric
  imperfections.
{\em International Journal of Non-Linear Mechanics},
  38(7):1119--1132, 2003.

\bibitem{Sharif-Khodaei:2008:MBM}
Z.~Sharif-Khodaei and J.~Zeman.
Microstructure-based modeling of elastic functionally graded
  materials: {O}ne dimensional case.
{\em Journal of Mechanics of Materials and Structures}, 2008.
accepted for publication,
  e-print:~\href{http://arxiv.org/abs/0802.0511}{arXiv/0802.0511}.

\bibitem{Spence:2008:PMS}
S.~M. Spence, M.~Gioffr\'{e}, and M.~D. Grigoriu.
Probabilistic models and simulation of irregular masonry walls.
{\em Journal of Engineering Mechanics ASCE}, 134(9):750--762, 2008.

\bibitem{SudDer00}
B.~Sudret and A.~Der~Kiureghian.
Stochastic finite element methods and reliability: {A} state of the
  art report.
Technical Report UCB/SEMM-2000/08, Univesity of California, Berkley,
  2000.
Available at
  \\\url{http://nisee.berkeley.edu/documents/SEMM/SEMM-2000-08.pdf} [\today].

\bibitem{Torq02}
S.~Torquato.
{\em Random heterogeneous materials: Microstructure and macroscopic
  properties}.
Springer-Verlag, 2002.

\bibitem{Vanm98}
E.~Vanmarke.
{\em Random fields: {A}nalysis and Synthesis}.
MIT Press, 1998.

\bibitem{Sejnoha:2004:HRMS}
M.~\v{S}ejnoha, J.~Zeman, and J.~Nov\'{a}k.
Homogenization of random masonry structures - {C}omparison of
  numerical methods.
In {\em EM 2004 - 17th ASCE Engineering Mechanics Division
  Conference}, page 8 pp., Newark, US, 2004. University of Delaware.
\newblock
  \\\url{http://chinacat.coastal.udel.edu/~kirby/EM2004/paperfinal/48.pdf}
  [\today].

\bibitem{Willis:1977:BSC}
J.~R. Willis.
Bounds and self-consistent estimates for the overall properties of
  anisotropic composites.
{\em Journal of the Mechanics and Physics of Solids}, 25:185--202,
  1977.

\bibitem{Xu:2007:MSFEM}
F.~X. Xu.
A multiscale stochastic finite element method on elliptic problems
  involving uncertainties.
{\em Computer Methods in Applied Mechanics and Engineering},
  196(25--28):2723--2736, 2007.

\bibitem{Xu:2005:SCM}
F.~X. Xu and L.~Graham-Brady.
A stochastic computational method for evaluation of global and local
  behavior of random elastic media.
{\em Computer Methods in Applied Mechanics and Engineering},
  194(42--44):4362--4385, 2005.

\bibitem{Xu:2006:CSH}
F.~X. Xu and L.~Graham-Brady.
Computational stochastic homogenization of random media elliptic
  problems using {F}ourier {G}alerkin method.
{\em Finite Elements in Analysis and Design}, 42(7):613--622, 2006.

\bibitem{Zeman:2007:FRM}
J.~Zeman and M.~\v{S}ejnoha.
From random microstructures to representative volume elements.
{\em Modelling and Simulation in Materials Science and Engineering},
  15(4):S325--S335, 2007.

\bibitem{ZhaHel94}
J.~Zhang and B.~Hellingwood.
Orthogonal series expansion of random fields in reliability analysis.
{\em Journal of Engineering Mechanics ASCE}, 120(12):2660--2677,
  1994.

\end{thebibliography}
\end{document}